\documentstyle[prc,aps,preprint]{revtex}
\tightenlines
\def\btau{\pmb{$\tau$}}

\def\be{\begin{equation}}
\def\ee{\end{equation}}
\def\bea{\begin{eqnarray}}
\def\eea{\end{eqnarray}}

\begin{document}

\begin{center}
{\large
Asymmetric nuclear matter in a Hartree-Fock approach to non-linear QHD}\\
[2ex]

V.Greco${}^{{1}}$, M.Colonna${}^{{1}}$,
M.Di Toro${}^{{1}}$, G.Fabbri${}^{{2}}$
and F.Matera${}^{{2}}$\\[2ex]

$^{1}$Laboratori Nazionali del Sud, Via S. Sofia 44,
I-95123 Catania, Italy\\
and Universit\`a degli Studi di Catania\\

$^{2}$Dipartimento di Fisica, Universit\`a degli Studi di Firenze\\
and INFN, Sezione di
Firenze,\\
L.go E. Fermi 2, I-50125, Firenze, Italy\\[2ex]
\end{center}
\begin{abstract}

The Equation of State ($EOS$) for asymmetric nuclear matter
is discussed starting from a
phenomenological hadronic field theory of Serot-Walecka type including
exchange terms. In a model with self interactions of the scalar
sigma-meson we show that the Fock terms naturally lead to
isospin effects in
the nuclear $EOS$. These effects are quite large and
dominate over the contribution due to
isovector mesons. 
We obtain a potential symmetry term of "stiff" type, i.e.
increasing with baryon density and an interesting behaviour of
neutron/proton effective masses of relevance for transport properties
of asymmetric dense matter.
\end{abstract}

\vspace{1cm}
\noindent
Phenomenological hadronic field theories (Quantum Hadrodynamics, $QHD$)
are widely used in dense nuclear matter studies since relativistic
effects are expected to increase with baryon density \cite{sewa86}.
In most of the previous works on the subject, the Relativistic Mean
Field ($RMF$) approximation of $QHD$ has been followed. In the $RMF$
the meson
fields are treated as classical fields and consequently 
a Hartree reduction of one body density matrices is used.

Although the model has driven
a large amount of results on relativistic effects in nuclear structure
and dynamics \cite{snr94,magi97,suto94,flw95,yst98}, the lack of
exchange terms has implied some non satisfying features of the theory
and some efforts have been done to try to cure this problem
\cite{Hor83,Fu89,bou87,berna}.
In the RMF theory
each meson field is introduced, with appropriated readjusted couplings,
just to describe the dynamics of a corresponding degree of freedom,
without mixing due to many-body effects. Neutral $\sigma$ and $\omega$
mesons are in charge of saturation properties, isospin effects are carried
by isovector $\delta$ [$a_0(980)$] and $\rho$ mesons and finally
spin terms are coming
from pseudoscalar $\pi$ and $\eta$ fields. In a sense the model represents
a straightforward extension of the One-Boson-Exchange ($OBE$) description of
nucleon-nucleon scattering.

The aim of this letter is to introduce explicit many-body effects
just evaluating exchange term contributions. We will get qualitative
new features of equilibrium ($EOS$) and dynamical properties of
asymmetric nuclear matter. In particular a new density dependence 
of the symmetry term is expected, at variance with the simple
linear increase predicted by the $\rho-$exchange mechanism in
the Hartree scheme.

As we already know from
non-relativistic effective interactions, like the Skyrme forces, Fock
terms play an essential role in symmetry breakings and consequent mixing
of different degrees of freedom. Similar effects are expected here.
In particular, in the context of the QHD model, essential properties
of nuclear matter come mostly from the two neutral strong meson fields. Hence
it is important to evaluate the Fock contribution associated  with these
fields.

We will focus our attention on isospin contributions to the nuclear $EOS$,
symmetry term and neutron/proton effective masses, and on the
relativistic transport equation for asymmetric nuclear matter.

We start from a {\it $QHD-II$} model \cite{sewa86}
where the nucleons
are coupled to neutral scalar $\sigma$ and vector $\omega$ mesons
and to the isovector $\rho$ meson.
Self-interaction terms of the $\sigma$-field were
originally introduced for renormalization reasons \cite{bobo77,bopr89}
and can also be considered as a way to parametrize the density dependence
of the $NN$ force.
Actually they are also describing medium effects essential to reproduce
important properties
(compressibility and nucleon effective mass) of nuclear matter
around saturation density.
The Lagrangian density for this model is given by:
\begin{eqnarray}
{\cal L} = {\bar {\psi}}[\gamma_\mu(i{\partial^\mu}-{g_V}{\cal V}^\mu
 - g_{\rho}{\bf {\cal B}}^\mu \cdot {\btau} )-
(M-{g_S}\phi)]\psi + {1 \over 2}({\partial_\mu}\phi{\partial^\mu}\phi
- {m_S}^2 \phi^2) \nonumber \\
- {a \over 3} \phi^3 - {b \over 4} \phi^4
- {1 \over 4} W_{\mu\nu}
W^{\mu\nu} + {1 \over 2} {m_V}^2 {\cal V}_\nu {{\cal V}^\nu}
- {1 \over 4} {\bf L}_{\mu\nu}\cdot
{\bf L}^{\mu\nu} + {1 \over 2} {m_{\rho}}^2 {\bf {\cal B}}_\nu\cdot
{{\bf {\cal B}}^\nu}
\end{eqnarray}


where
$W^{\mu\nu}(x)={\partial^\mu}{{\cal V}^\nu}(x)-
{\partial^\nu}{{\cal V}^\mu}(x)~$ and
${\bf L}^{\mu\nu}(x)={\partial^\mu}{{\bf {\cal B}}^\nu}(x)-
{\partial^\nu}{{\bf {\cal B}}^\mu}(x)~.$

Here $\psi(x)$ generally denotes the
fermionic field, $\phi(x)$ and ${{\cal V}^\nu}(x)$ represent neutral scalar
 and
vector boson fields, respectively. ${{\bf {\cal B}}^\nu}(x)$
is the charged vector
field and ${\btau}$ denotes the isospin matrices. 

From previous equation one can derive the field equations and the
canonical energy-momentum tensor \cite{sewa86}.

In our approach we will perform the many-body calculations in the
quantum phase space introducing the Wigner transform of the
one-body density matrix for the fermion field. This method has
two main advantages, the use of physical quantities
and the
direct derivation of dynamical transport equations.
The one--particle Wigner function is defined as:
$$[{\widehat F}(x,p)]_{\alpha\beta}=
{1\over(2\pi)^4}\int d^4Re^{-ip\cdot R}
<:\bar{\psi}_\beta(x+{R\over2})\psi_\alpha(x-{R\over2}):>~,$$
here $\alpha$ and $\beta$ are indices for intrinsic degrees of freedom
of the fermionic field (spin and isospin).
The brackets denote statistical averaging and the double dots denote
normal ordering.
The equation of motion for the Wigner function can be derived from
the Dirac field equation by using standard procedures
(see e.g. Refs.\cite{degr80,hak78}),
it reads:
\begin{eqnarray}
{i\over2}{\partial_\mu}[\gamma^\mu{\hat F}(x,p)]_
{\alpha\beta}+
p_\mu[{\gamma^\mu}{\hat F}(x,p)]_{\alpha\beta}-
M{\hat F}_{\alpha\beta}(x,p) \nonumber \\
-g_V{1\over(2\pi)^4}\int d^4Re^{-ip\cdot R}
<:\bar{\psi}_\beta(x_+){\gamma^\mu_{\alpha\delta}}\psi_\delta(x_-)
{\cal V}_\mu(x_-):> \nonumber \\
+g_S{1\over(2\pi)^4}\int d^4Re^{-ip\cdot R}
<:\bar{\psi}_\beta(x_+)\psi_\alpha(x_-)\phi(x_-):> \nonumber \\
-g_{\rho}{1\over(2\pi)^4}\int d^4Re^{-ip\cdot R}
<:\bar{\psi}_\beta(x_+){\gamma^\mu_{\alpha\delta}}\psi_\delta(x_-)
{\btau}\cdot {\bf{\cal B}}_\mu(x_-):> =0~,
\end{eqnarray}
with $x_+=x+{R\over2}$ and $x_-=x-{R\over2}$.
\par

In order to take into account the contribution of exchange terms in a
manageable way we assume, as a
basic approximation, that in the equations of motion for the meson
fields the terms containing
derivatives can be neglected with
respect to the mass terms. Therefore the meson  
field operators are directly connected to the operators of
the nucleon scalar and current densities:
\begin{eqnarray}
{\widehat{\Phi}/f_S} + A{\widehat{\Phi^2}}
+ B{\widehat{\Phi^3}}&=\bar\psi(x)\psi(x)~,\nonumber \\
{\widehat V}^\mu(x)&=f_V\bar\psi(x){\gamma^\mu}\psi(x)~,\nonumber \\
{\widehat  {\bf B}}^\mu(x)&=f_{\rho}\bar\psi(x){\gamma^\mu} {\btau}
\psi(x)~,
\end{eqnarray}
where $f_S = (g_S/m_S)^2$, $f_V = (g_V/m_V)^2$, $f_{\rho} = (g_{\rho}/
m_{\rho})^2$ and
${\widehat \Phi} = g_S \phi$, ${\widehat V^{\mu}} = g_V {\cal V}^{\mu}$,
$A = a/g_S^3$, $B = b/g_S^4$, ${\bf B}^\mu = g_{\rho} {\bf \cal{B}}^\mu$.
After substituting in Eq.(2) these expressions for the meson
field operators, we obtain an equation which contains only
nucleon field operators.\par
The present approximation implies that retardation and finite size
effects in the exchange of mesons between nucleons are
neglected. However, we are concerned with a semiclassical
description of nuclear dynamics, so that the nuclear medium
is supposed to be in states for which the nucleon scalar and
current densities are smooth functions of the space-time
coordinates. Therefore, because of the small Compton wave--lengths
of the mesons $\sigma$, $\omega$ and $\rho$, the assumptions
expressed by Eq.s(3) are quite reasonable.
For light mesons such as pions this approximation is not justified.
The inclusion of self--interaction terms of the pionic field
requires a different approximation scheme. However, it
has been shown that the inclusion of pions does not change
qualitatively the description of nuclear matter but, as expected,
 in very low density regions \cite{Hor83}.\par
An attempt to include exchange terms in the $QHD$ approach was
previously performed without self--interaction terms for the
$\sigma$ field \cite{bou87}, with results not satisfying
due to the inadequacy of the model.
Within the same model, a
relativistic kinetic equation with self--consistent mean field
has been derived in Ref.\cite{dema91} taking into account the
exchange terms. Here
we evaluate the effects in a more physical model with self-interacting
higher order $\sigma$ terms. \par
In order to evaluate the nuclear Equation of State
the quantity of interest is the statistical
average of the canonical energy-momentum density tensor.
According to our approximation, where 
terms containing the derivatives of the meson fields are neglected,
the tensor takes the form:
\begin{equation}
T_{\mu\nu}(x) = {i\over 2}{\bar \psi}(x)\gamma_\mu \partial_{\nu} \psi(x)
+ [ U({\widehat \Phi})
 - {1\over 2}{\widehat V}_\lambda {\widehat V}^\lambda
/f_V
- {1\over 2}{\widehat {\bf B}}_\lambda\cdot {\widehat {\bf B}}^\lambda
/f_\rho]g_{\mu\nu},
\end{equation}
where $g_{\mu\nu}$ is the diagonal metric tensor and
$U({\widehat \Phi}) =
{1\over 2}{\widehat \Phi}^2/f_S + A/3~ {\widehat \Phi}^3 + B/4~
{\widehat \Phi}^4$.
Following the treatment of the Fock terms in non--linear QHD  
introduced in Ref.\cite{prc_nuovo},
the energy-momentum density tensor is given by:
\begin{eqnarray}
<T_{\mu\nu}(x)> =  8\int d^4 p~ p_{\nu}F_{\mu}(x,p) +
\{U(\Phi) - f_V/2~ j_\lambda
j^\lambda\ - f_{\rho}/2~ b_\lambda b^\lambda\}g_{\mu\nu} \nonumber \\
- {1\over 2}[ 
Tr {\widehat F}^{2}(x)
{{d^{2}U(\Phi)}\over {d\rho_S^{2}}}
- f_V Tr (\gamma_\lambda {\widehat F}(x)
\gamma^\lambda {\widehat F}(x))
- f_{\rho} Tr (\gamma_\lambda {\btau}\cdot {\widehat F}(x)
\gamma^\lambda {\btau} {\widehat F}(x))]g_{\mu\nu}\, ,
\end{eqnarray}
where $F_{\mu}(x,p)$ is the isoscalar vector component of the Wigner
function and the classical value of the $\sigma$ field, $\Phi(x)$,
obeys the equation $\Phi/f_S + A~{\Phi}^2 +
B~{\Phi}^3=\rho_S(x)=\,<:{\bar \psi}(x) \psi(x):>$.  
The matrix ${\widehat F}(x)=\,\int d^4 p~ {\widehat F}(x,p)$
can be decomposed in components with
definite transformation properties (Clifford algebra),
where, for instance, the scalar and the vector components of the isoscalar
part ($F$,$F_{\mu}$) are related to scalar and current densities as:
$8\,F(x) = \rho_S(x)$, 
$8\,F^{\mu}(x) = j^{\mu}(x) = 
<:{\bar \psi}(x) \gamma^{\mu} \psi(x):>$. The corresponding
components of the isovector part are related to
isovector densities: $b(x) = <:{\bar \psi}(x)\tau_3 \psi(x):>$
(~scalar~), $b^{\lambda}(x) =
<:{\bar \psi}(x) \tau_3 \gamma^{\lambda} \psi(x):>$ (~vector~).
\par
The quantities in square brackets of Eq.(5)
are the contributions of the exchange terms.
It is essential to note that Fock terms contain traces of powers of
${\widehat F}(x)$
that naturally bring scalar, vector, tensor, pseudoscalar and pseudovector
contributions. In particular for the case of asymmetric nuclear matter
we obtain scalar and vector isovector contributions to the $EOS$, generally
associated respectively with $\delta$ and $\rho$ mesons.
From Eq.(6) we obtain the energy density for asymmetric nuclear
matter that in analogy to the Hartree case can be rewritten
in the following form:

\begin{equation}
\epsilon= {<T_{00}>} = {\epsilon_{kin}}^p+{\epsilon_{kin}}^n+{U(\Phi)}+
{1\over2}
{\tilde f_S} {\rho_S}^2 + {1\over2} {\tilde f_V} {\rho_B}^2 + {1\over2} 
f'_S b^2+
{1\over2} f'_V {b_0}^2
\end{equation}
where $\rho_B$ is the baryon density and
$b_0={\rho}_{Bp} -{\rho}_{Bn}$
is the corresponding isovector density. The

\begin{equation}
{\epsilon_{kin}}^i={2\over{(2{\pi})^3}} \int d^3 p \sqrt{p^2+{{M_i}^*}^2}
= {1\over4} [3 {E_i}^*{\rho_{Bi}} + {M_i}^*{\rho_{Bi}}]~~~i=n,p
\end{equation}
are kinetic contributions and
$$
{\tilde{f_S}}={1\over2} f_V - {1\over8}({{d\Phi}\over{d\rho_S}} +
 \rho_S {{d^2\Phi}\over{d\rho_S^2}}) + {3\over2}~f_{\rho};
$$
$$
{f'_S}={1\over2} f_V - {1\over8}({{d\Phi}\over{d\rho_S}} +
 \rho_S {{d^2\Phi}\over{d\rho_S^2}}) - {1\over2}~f_{\rho};
$$
$$
{\tilde f_V}={5\over4} f_V + {1\over8}({{d\Phi}\over{d\rho_S}} -
 \rho_S {{d^2\Phi}\over{d\rho_S^2}}) + {3\over4}~f_{\rho};
$$
\begin{equation}
{f'}_V={1\over4} f_V + {1\over8}({{d\Phi}\over{d\rho_S}} -
 \rho_S {{d^2\Phi}\over{d\rho_S^2}}) + {3\over4}~f_{\rho}
\end{equation}
are density dependent effective coupling constants.
$E^*_i = \sqrt{|{\bf p}|^2 + {M^*_i}^2}$ and $M^*_i$ are
the effective masses, see Eq.s.(9,10).

Here we explicitly obtain
a density dependence arising also in the vector, isovector and isoscalar
couplings, like in the phenomenological approach of Ref.\cite{tywo99}.
We have verified that the used approximation leads to a thermodynamically
consistent theory \cite{prc_nuovo}.

We remark that, as in non-linear mean-field models,
we have in total five parameters. As usual, the ones related to isoscalar
mesons, $f_V,
f_S,A,B$ are fixed in order to reproduce equilibrium properties
of symmetric nuclear matter: saturation density $\rho_0=0.16~fm^{-3}$,
binding energy $E/A= -16~MeV$, compressibility  modulus $K_0 =245~MeV$
and nucleon effective (or Dirac) mass at $\rho_0$,
${M_0}^*=0.73~M$ (see Eq.(10)). This value for ${M_0}^*$ is in the
range expected from the analysis of elliptic--flow data
\cite{Dan00}. The coupling constant $f_\rho$ can then
be adjusted in order to get a
good value for the symmetry energy at saturation density, but now taking into
account the contribution to the isovector channel coming from the isoscalar
mesons through the Fock terms. In our calculations we have a symmetry
coefficient of the Weiszaecker mass formula $a_4 = 31.5~MeV$.
In the Table we list the values obtained for the five parameters.

\begin{center}
\begin{tabular}{|c|c|c|c|c|}
\hline
${f_V}$ & ${f_S}$ &  $A$ & $B$ & ${f_{\rho}}$ \\
$(fm^2)$ & $(fm^2)$ & $(fm^{-1})$ &  &$(fm^2)$\\
\hline
3.998  &  9.731  &  0.088  &  -0.015 & 0.6  \\
\hline
\end{tabular}
\end{center}
\begin{quotation}
{\bf Table} $NLHF$ parameters from the fit to saturation properties
of nuclear matter (see the text).
\end{quotation}

According to these values, we see that the dominant contributions
to the density dependent coupling functions ${\tilde f_S}$, ${\tilde f_V}$,
${\tilde f'_S}$, ${\tilde f'_V}$, Eq.(8), come essentialy from the isoscalar
$\sigma$ and $\omega$ mesons.

We discuss now some results for the $EOS$ of asymmetric nuclear matter.
We show the comparison between our {\it Non Linear Hartree-Fock}
($NLHF$) present calculations and those of the  {\it Non Linear Hartree}
($NLH$) model of
Ref.\cite{sewa86,bopr89}, including the isovector $\rho$ and $\delta$
mesons \cite{kuku97},
with parameters chosen in order to give the same
saturation properties.

The comparison for potential symmetry energies per nucleon is presented
in Fig.1.
As already stressed, in the $NLHF$
results a large contribution to the symmetry term
comes from the Fock contributions associated with the $\sigma$ and 
$\omega$ mesons,
with the corresponding four parameters of the theory fitted on
properties of symmetric nuclear matter.
Therefore the
inclusion of the Fock terms (solid line, $NLHF$) can give the correct value
of $a_4$ even with a relatively small coupling constant for the $\rho$ meson:
$f_{\rho} = 0.6~fm^2$, close to the free space value. 
We remind that this large exchange
contribution to the symmetry energy occurs
also in $QHD$ without non--linear terms \cite{bou87}.\par
We show also $NLH$ calculations including both $\rho$ and the isovector
scalar $\delta$ mesons (dashed line).  In this case  the coupling constants
$f_\rho$ and $f_\delta$ have been chosen in order to reproduce
at saturation density the same symmetry energy
and neutron-proton effective mass splitting that we
get within our model. We remark that now we need a $f_\rho=2.3~fm^2$, {\it
about four times the free space value}, and also a relatively strong $\delta$
coupling, $f_\delta=1.4~fm^2$, but still in the range of free space values
\cite{kuku97,mac89}.
For reference we show also the result of a $NLH$ calculations including
{\it only} the $\rho$ contribution (long-dashed line). In order to have
the same $a_4$ value at saturation density we need a $f_\rho=1.2~fm^2$,
still almost two times the free space value. We stress that the inclusion of
the $\delta$ contribution in the Hartree scheme, necessary for the
neutron-proton mass splitting, gives also an attractive term in the symmetry
energy, see Ref.\cite{kuku97}, and so a much stronger $\rho$ coupling is
required in order to reproduce the correct $a_4$ coefficient around
saturation density.

In all these relativistic models a quite repulsive density dependence
of the symmetry term of the $EOS$ is obtained.

We notice that the density dependence of the symmetry energy that one
obtains in the complete $NLH+\rho+\delta$ model is quite different with
respect to our results.
This is due to the fact that
in the $NLHF$ model the coupling functions in the isovector channels
($f'_S$, $f'_V$) become density dependent. This represents a 
{\it qualitative new effect of the exchange terms in a non-linear
$QHD$ model}.

Such density dependence is shown in Figure 2. In particular we stress the
behaviour at sub-nuclear densities due to the opposite sign of the
$d\Phi/d\rho_S$ contribution, see Eq(9). This implies a "softer"
behaviour of the potential symmetry term {\it below saturation density}
in the $NHLF$ case [see the insert in Fig.1].

In Fig.3 we report the density dependence of neutron (bottom) and
proton (top) effective masses for various asymmetries ($I=(N-Z)/A$)
as predicted by $NLHF$ (solid lines) and $NLH+\rho+\delta$ (dashed lines).
We remind that in the usual Hartree approximation
this effect is associated with the scalar isovector $\delta$ meson. 
 The Fock terms lead to a behaviour:
\begin{equation}
M^*_{n,p}(NLHF) = M^* \pm f^m_S\,b +
{{b^2+b_0^2} \over 16} {{d^2\Phi}\over{d{\rho_S^2}}} ~~~~(+\equiv p,
-\equiv n),
\end{equation}
where
$M^*$ is the nucleon effective mass in symmetric nuclear matter,
\begin{equation}
M^* = M - {\Phi} -(f^m_S +
2 f_\rho) \rho_S
+ {1\over 16} (\rho_S^2 + \rho_B^2)
{{d^2\Phi}\over{d\rho_S^2}}.
\end{equation}
and
\begin{equation}
f^m_S = {f_V \over 2} - {1 \over 8}{{d\Phi}\over{d\rho_S}} -
{f_{\rho} \over 2}\, .
 \nonumber
\end{equation}
Since the coefficient $f^m_S $ is positive
we get an effect very similar to what expected from the contribution of the
$\delta$
meson \cite{kuku97}, dashed lines in Fig.3.
The splitting of proton and neutron effective masses influences
also the density behaviour of the symmetry energy (Fig.1) and
is responsible for its rapid increase at high density.
On this point we would like to make two more remarks:

i) The splitting is quite small around normal density,
so it can be neglected in finite nuclei. This difference
in $n$ and $p$ effective masses can however be
relevant for transport properties of asymmetric, dense
nuclear matter that can be formed in intermediate energy reactions
with radioactive beams, naturally apart neutron star properties;

ii) The proton effective masses are systematically above the neutron ones.
This trend, also in agreement with $\delta$ calculations \cite{kuku97,dele98},
is just the opposite of what expected from Brueckner-Hartree-Fock
calculations with realistic $NN$ potentials \cite{zbl99}. Although
relativistic and non-relativistic effective masses cannot be
directly compared, see ref.s \cite{mass,bou87}, it is interesting to look at
the predictions given by Skyrme--like effective forces. Calculations
based on Skyrme forces are, to some extent, the non--relativistic
counterpart of our approach, also because exchange contributions
are suitably accounted for.
Concerning the splitting of neutron
and proton effective masses in asymmetric matter, the most recent 
parametrizations, SLy-type 
\cite{Cha97}, of Skyrme forces give the proton effective mass
above the neutron one, in agreement with our calculations.
Previous parametrizations, instead, yield a splitting in the
opposite direction, but also show unpleasant behaviours in the spin channel
(collapse of polarized neutron matter, see discussion in \cite{Cha97}). 
In the insert of Fig.3 the ratio of the
splitting, $\delta M^*=\, M^*_p- M^*_n$, to the bare
nucleon mass is displayed for the SLy4 force and for our
approach.

The sign of the splitting depends
on the chosen effective interaction. This puzzle can be disentangled
by a detailed analysis of the transport properties of dense
asymmetric nuclear matter. 

In conclusion we have shown the
evaluation, {\it in a non-perturbative
scheme}, of Fock term contributions in a non linear effective field
theory for asymmetric nuclear matter. Very reasonable and stimulating results
for isospin effects on the nuclear $EOS$ are obtained just
from such minimal explicit many-body effects.

We stress again that at variance with non relativistic effective
forces all the $RMF$ models give a {\it stiff} potential symmetry term,
i.e. more repulsive with increasing baryon density. Moreover
the density dependence of the isovector couplings due to the Fock
contributions leads to a new interesting effect (see the $NLHF$ curve
in the insert of Fig.1), a {\it softening} of the behaviour
at sub-nuclear densities. We expect to see related dynamical effects
in heavy ion collisions at intermediate energies in fragmentation
events \cite{cdl98} and collective flows \cite{scd99,bao00}.

 Of course a similar analysis can be
performed for spin effects. Moreover a transport equation can
be consistently derived to be used for the study of dynamical evolution of
nuclear matter far from normal conditions.

\newpage

\begin{center}
{FIGURE CAPTIONS}
\end{center}

\noindent
{\bf Figure 1 - }
Potential symmetry energy per nucleon vs. baryon density
(in units of saturation density).
Solid line: $NLHF$ results. Dashed line: Hartree results with
$\rho$ and $\delta$ mesons ($NLH$). Long-dashed: Hartree results with
only $\rho$ meson. 

\vskip 1.0cm
\noindent
{\bf Figure 2 - }
Density dependence of $f'_S$ and $f'_V$. Each curve is normalized
to the value at saturation density.

\vskip 1.0cm
\noindent
{\bf Figure 3 - }
Proton (top curves) and neutron (bottom curves)
effective masses vs. baryon density for various charge
asymmetries. $I={{N-Z}\over{N+Z}}=0$ : long dashed line.
$I=0.8$ : solid lines $NLHF$; dashed lines $NLH$.
In the insert, the relative splitting of neutron and proton
effective masses. Solid line: NLHF results, circles: SLy4 \cite{Cha97}
non--relativistic results.    

\vskip 0.6truecm
\begin{thebibliography}{99}

\bibitem{sewa86} B.D. Serot and J.D. Walecka, Adv. Nucl. Phys. 16 (1986) 1,\\
B.D. Serot and J.D. Walecka, Int. J. Mod. Phys. E 6 (1997) 515.
\bibitem{snr94} M.M. Sharma, M.A. Nagarajan and P. Ring, Ann. Phys. (NY)
231 (1994) 110,\\
P. Ring, Progr. Part. Nucl. Phys. 37 (1996) 193.
\bibitem{magi97} Z.Y. Ma , N. Van Giai, H. Toki and M. L'Huillier,
Phys. Rev. C 55 (1997) 2385 and Ref.s therein.
\bibitem{suto94} Y. Sugahara and H. Toki, Nucl. Phys. A 579 (1994) 557.
\bibitem{flw95} Ch. Fuchs, H. Lenske and H.H. Wolter,
Phys. Rev. C 52 (1995) 3043.
\bibitem{yst98} S. Yoshida, H. Sagawa and N. Tagikawa,
 Phys. Rev. C 58 (1998) 2796.
\bibitem{Hor83}C.J. Horowitz and B.D. Serot, Nucl.Phys. A 399 (1983) 529.
\bibitem{Fu89} R.J. Furnstahl, R.J. Perry, and B.D. Serot,
Phys. Rev. C 40 (1989) 321.
\bibitem{bou87} A. Bouyssy, J.F. Mathiot, N. Van Giai and S. Marcos,
Phys.Rev. C 36 (1987) 380, \\M. Lopez-Quelle, S. Marcos, R. Niembro,
A. Bouyssy and N. Van Giai, Nucl. Phys. A 483 (1988) 479.
\bibitem{berna} P. Bernardos, V.N. Fomenko, N. Van Giai, M. Lopez-Quelle,
S. Marcos, R. Niembro and L.N. Savushkin, Phys. Rev. C 48
(1993) 2665,\\ J. Ramksch\"utz, F. Weber and M.K. Weigel,
J. Phys. G 16 (1990) 987.
\bibitem{bobo77} J. Boguta and A.R. Bodmer, Nucl. Phys. A 292 (1977) 413.
\bibitem{bopr89} J. Boguta and C.E. Price, Nucl. Phys. A 505 (1989) 123.
\bibitem{degr80} S.R. de Groot, W.A. van Lee and Ch.G. van Weert,
{\it Relativistic Kinetic Theory}, North-Holland Amsterdam 1980.
\bibitem{hak78} R. Hakim, Nuovo Cimento 6 (1978) 1.
\bibitem{dema91} A. Dellafiore and F. Matera, Phys. Rev. C 44 (1991) 2456.
\bibitem{prc_nuovo} V. Greco, F. Matera, M. Colonna, M. Di Toro and G. Fabbri,
Phys. Rev. C 63 (2001) 035202.
\bibitem{tywo99} S. Typel and H.H. Wolter, Nucl. Phys. A 656 (1999) 331.
\bibitem{Dan00} P. Danielewicz, Nucl. Phys. A 673 (2000) 375. 
\bibitem{kuku97} S. Kubis and M. Kutschera, Phys. Lett. B399 (1997) 191.
\bibitem{mac89} R. Machleidt, Adv. Nucl. Phys. 19 (1989) 189,\\
R. Brockmann and R. Machleidt, Int. Rev. Nucl. Phys.
Vol.8, Ch.2, M. Baldo Ed., World Sci. 1999.
\bibitem{dele98} F. de Jong and H. Lenske, Phys. Rev. C 58 (1998) 890.
\bibitem{mass} A.Bouyssy and S.Marcos, Phys. Lett. B 127 (1983) 157
\bibitem{zbl99} W. Zuo, I. Bombaci and U. Lombardo, Phys.Rev. C 60
(1999) 24605 and Ref.s therein.
\bibitem{Cha97} E. Chabanat, P. Bonche, P. Haensel, J. Meyer and
R. Schaeffer, Nucl. Phys. A 627 (1997) 7B10. 
\bibitem{cdl98} M. Colonna, M. Di Toro and A.B. Larionov, Phys. Lett. B 428
 (1998) 1.
\bibitem{scd99} L. Scalone, M .Colonna and M. Di Toro, Phys. Lett. B 461
 (1999) 9.
\bibitem{bao00} Bao-An Li, Phys. Rev. Lett. 85 (2000) 4221.
\end{thebibliography}
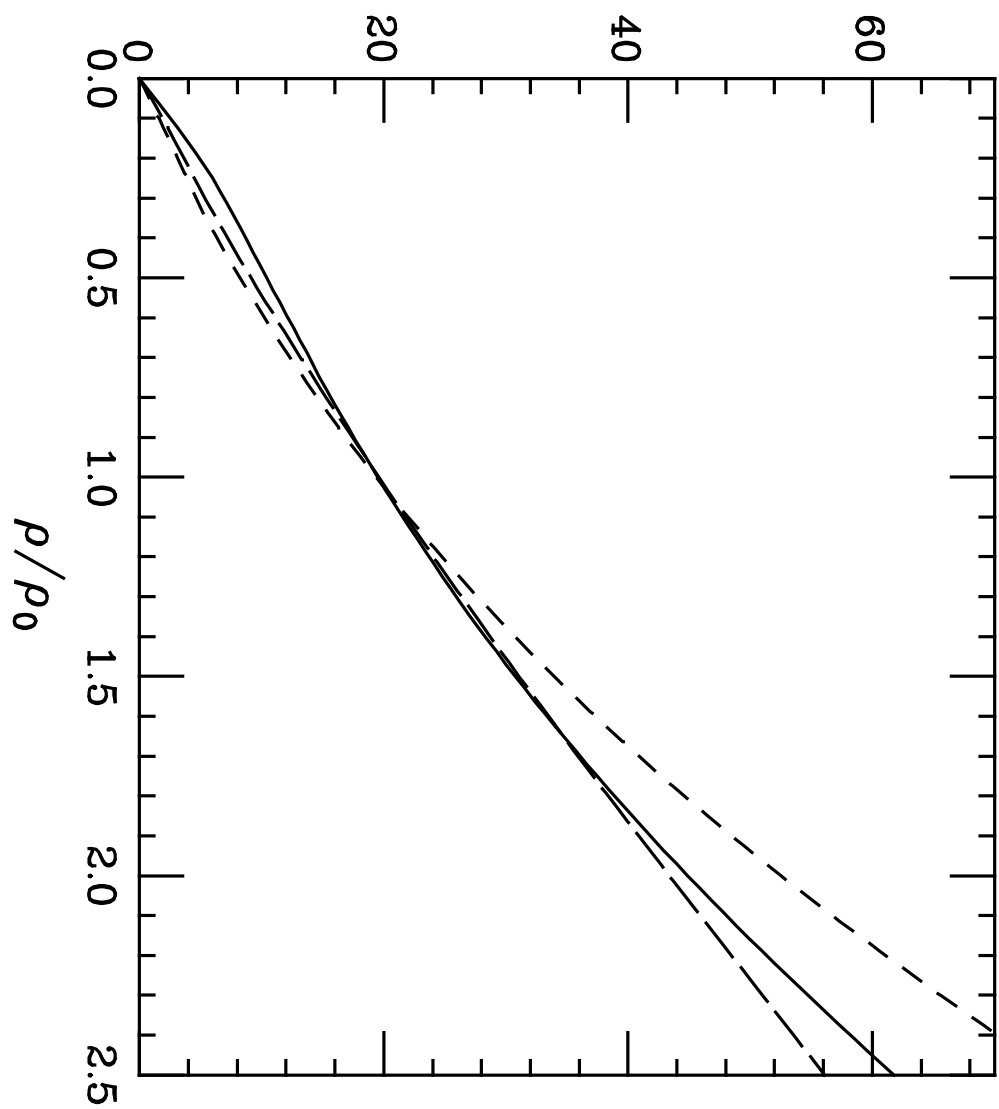
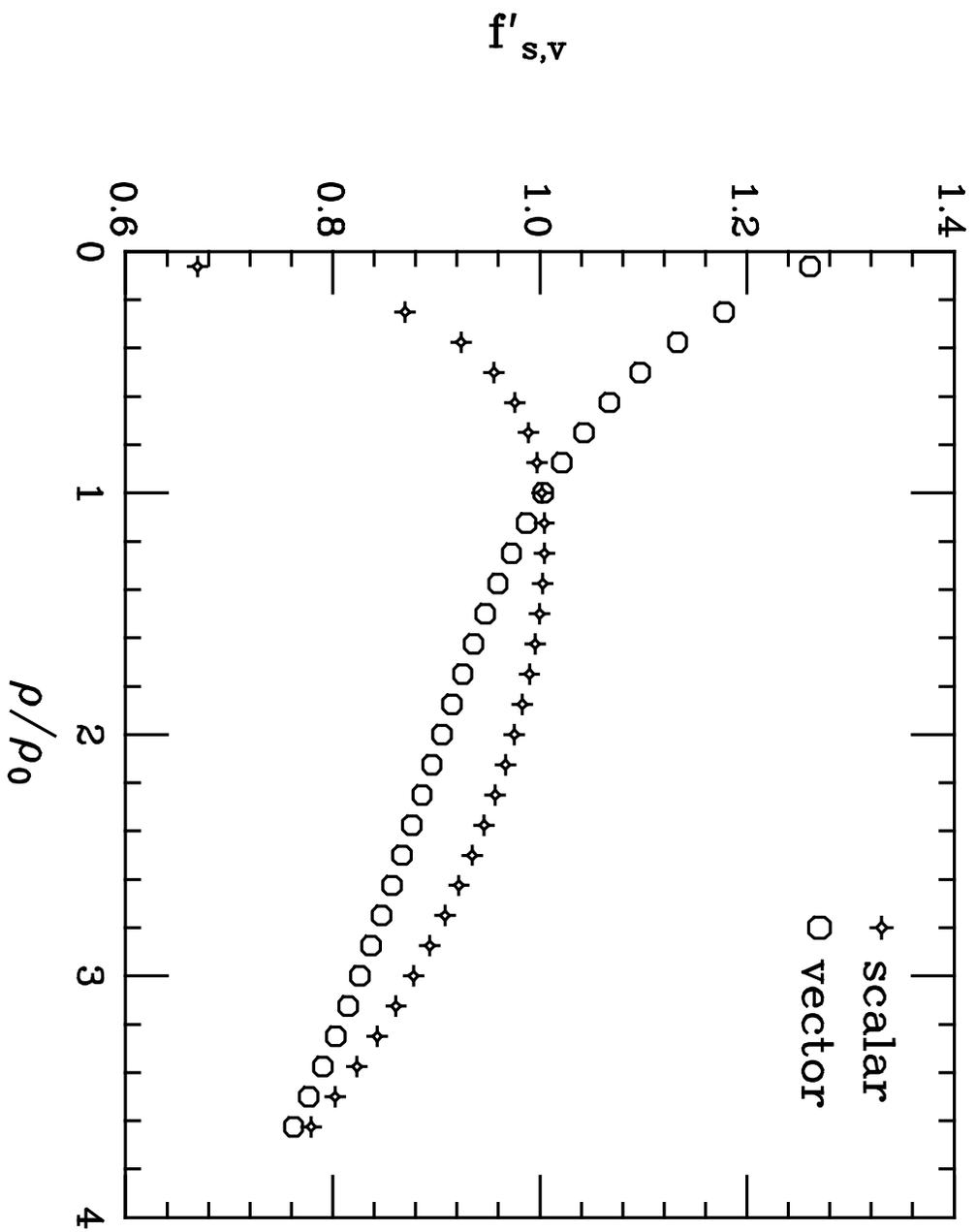
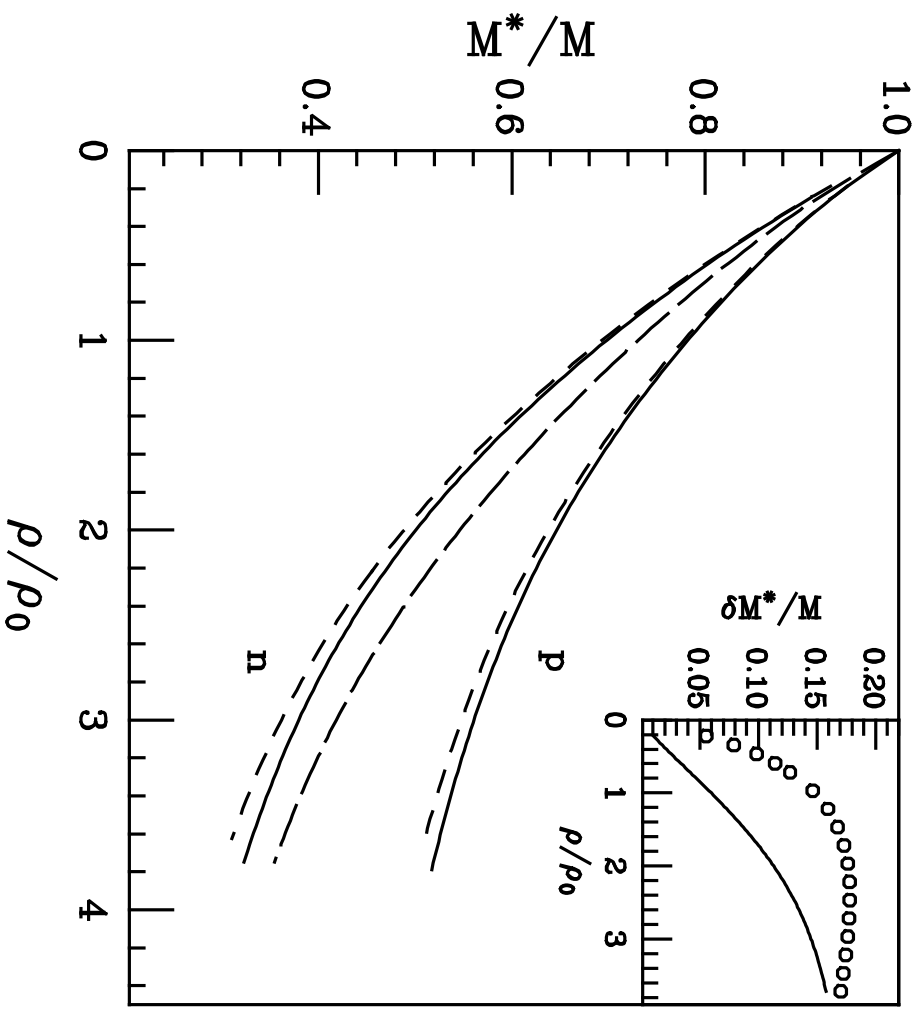
\end{document}